\documentclass[11pt, a4paper]{article}

\usepackage[dvipdfmx]{graphicx}
\usepackage{amssymb}
\usepackage{amsmath}
\usepackage{color}
\usepackage{subcaption}
\usepackage[sort&compress,numbers, merge]{natbib}
\usepackage[colorlinks=true, linkcolor=OliveGreen, citecolor=RoyalBlue, urlcolor=Brown]{hyperref}

\definecolor{Orange}{cmyk}{0,0.61,0.87,0}
\definecolor{JungleGreen}{cmyk}{0.99,0,0.52,0}
\definecolor{OliveGreen}{cmyk}{0.64,0,0.95,0.40}
\definecolor{Brown}{cmyk}{0,0.81,1,0.60}
\definecolor{RoyalBlue}{cmyk}{0.71,0.53,0,0.12}

\newcommand{\lsim}{\raise0.3ex\hbox{$\;<$\kern-0.75em\raise-1.1ex\hbox{$\sim\;$}}}
\newcommand{\gsim}{\raise0.3ex\hbox{$\;>$\kern-0.75em\raise-1.1ex\hbox{$\sim\;$}}}

\newcommand{\bmx}{\left(\begin{array}}
\newcommand{\emx}{\end{array}\right)}

\newcommand{\del}[2]{\frac{\partial #1}{\partial #2}}

\newcommand{\lmlt}{U(1)$_{L_\mu - L_\tau}$ }

\begin{document}

\begin{titlepage}

\begin{flushright}
{\tt 
STUPP-20-242\\
UME-PP-18
}
\end{flushright}

\vskip 1.35cm
\begin{center}

{\Large
{\bf
Resolving the Hubble tension in a U(1)$_{L_\mu-L_\tau}$ model with Majoron
}
}

\vskip 1.2cm

Takeshi Araki$^{1}$,
Kento Asai$^{2}$, 
Kei Honda$^{3}$, 
Ryuta Kasuya$^{3}$, 
Joe Sato$^{3}$, 
Takashi Shimomura$^{4}$, 
Masaki J.S. Yang$^{3}$

\vskip 0.4cm

{\it ${}^1$Faculty of Dentistry, Ohu University, 31-1 Sankakudo, Tomita-machi, Koriyama, Fukushima 963--8611, Japan} \\[2pt]
{\it ${}^2$Department of Physics, University of Tokyo, Bunkyo-ku, Tokyo 133--0033, Japan} \\[2pt]
{\it ${}^3$Department of Physics, Saitama University, 255 Shimo-Okubo, Sakura-ku, Saitama 338--8570, Japan} \\[2pt]
{\it ${}^4$Faculty of Education, University of Miyazaki, 1-1 Gakuen-Kibanadai-Nishi, Miyazaki 889--2192, Japan}

\date{\today}

\vskip 1.5cm

{\bf {\Large Abstract}}
\end{center}

In this paper, we explore possibilities of resolving the Hubble tension and $(g-2)_{\mu}$ anomaly simultaneously in a \lmlt model with Majoron.
We only focus on a case where the Majoron $\phi$ does not exist at the beginning of the universe and is created by neutrino inverse decay $\nu\nu\to \phi$ after electron-positron annihilation. %
In this case, contributions of the new gauge boson $Z'$ and Majoron $\phi$ to the effective number of neutrino species $N_{\rm eff}$ can be calculated in separate periods. 
These contribution are labelled $N'_{\rm eff}$ for the \lmlt gauge boson and $\Delta N_{\rm eff}^\prime$ for the Majoron.
The effective number $N_{\rm eff} = N'_{\rm eff} + \Delta N_{\rm eff}^\prime$ is evaluated by the evolution equations of the temperatures and the chemical potentials of light particles in each period.

As a result, we found that the heavier $Z'$ mass  $m_{Z^\prime}$ results in the smaller $N_{\mathrm{eff}}^\prime$ and  requires the larger $\Delta N_{\mathrm{eff}}^\prime$ to resolve the Hubble tension. 
Therefore, compared to previous studies, the parameter region where the Hubble tension can be resolved is slightly shifted toward the larger value of $m_{Z^\prime}$.

\end{titlepage}

\section{Introduction}
\label{sec:introduction}

Recently, a discrepancy has been reported on the values of the Hubble constant $H_0$ from the cosmic microwave background (CMB) measurements~\cite{Aghanim:2019ame} and local measurements~\cite{Riess:2018uxu,Riess:2018byc,Riess:2019cxk,Wong:2019kwg}.
The inferred value from the $\Lambda$CDM with the temperature anisotropy of the CMB measured by Planck~\cite{Aghanim:2019ame}  is $H_0 = 67.36 \pm 0.54$ km/s/Mpc. On the other hand, the local measurements using Cepheids~\cite{Riess:2018uxu,Riess:2018byc} and 
type-Ia supernovae~\cite{Riess:2019cxk} by SH0ES reported larger values as $H_0 = 73.45 \pm 1.66$ km/s/Mpc and $74.03 \pm 1.42$ km/s/Mpc, respectively. 
A similar value of $H_0$ has been also reported by H0LiCOW from gravitational lensing with late time~\cite{Wong:2019kwg}. 
These local measurements result in a larger value of $H_0$ than the CMB measurement\footnote{
Local measurements based on the TRGB method~\cite{Freedman:2019jwv} and TDCOSMO+SLACS analyses~\cite{Birrer:2020tax} reported consistent values to the CMB results. 
}. 
The discrepancy reaches the level of $4-6\sigma$ and is called the Hubble tension.

Although the tension could originate from systematic errors in the measurements~\cite{Efstathiou:2013via,Freedman:2017yms,Ivanov:2020mfr}, it would indicate modifications of the standard cosmological model. 
Then, several solutions have been proposed in the fields of cosmology and particle physics. 
One of the approaches to solve the tension is to modify the effective number of neutrino species $N_{\mathrm{eff}}$. 
In Ref.~\cite{Aghanim:2019ame}, combining the results from the CMB, Cepheids and others, $N_{\mathrm{eff}}$ is derived as $3.27 \pm 0.15$ at $68\%$ C.L.~\cite{Aghanim:2018eyx}, which implies the difference from the $\Lambda$CDM results as $0.2 \lesssim \Delta N_{\mathrm{eff}} \lesssim 0.5$ to ameliorate the Hubble tension.~\footnote{
We should note that increasing $N_{\rm eff}$ worsens another milder tension relative to $\sigma_8$~\cite{Planck:2018vyg,Hildebrandt:2018yau} that is the cosmological parameter about the matter density fluctuation amplitude on 8 Mpc scales.}
Such a difference can be obtained when new interactions with neutrinos exist. 
In this regard, models with gauged \lmlt symmetry are very interesting~\cite{Foot:1990mn,He:1990pn,Foot:1994vd,He:1991qd}, under which only mu and tau-type leptons are charged. 
It is well-known that the long-standing discrepancy of the muon anomalous magnetic moment, $(g-2)_\mu$, can be resolved by the contributions of the new gauge boson $Z'$ with an MeV scale mass~\cite{Gninenko:2001hx,Baek:2001kca,Ma:2001md}. 
The new interaction also alters the decoupling time of neutrinos from the thermal bath at the early universe. 
In particular, the decays of $Z'$ to heat neutrinos lead to the increase of $N_{\mathrm{eff}}$.
In Ref.~\cite{Escudero:2019gzq}, it was shown that the Hubble tension can be solved simultaneously with the discrepancy of $(g-2)_\mu$.

Other interesting models are the ones with global Lepton number symmetry U(1)$_L$. 
In the class of seesaw mechanism, tiny neutrino masses are explained by the heavy Majorana masses of right-handed neutrinos which often can be 
generated by the spontaneous breaking of the Lepton number symmetry.  
As a result, a pseudo Nambu-Goldstone boson, the so-called Majoron, appears in the spectrum~\cite{Chikashige:1980ui,Gelmini:1980re,Georgi:1981pg,Schechter:1981cv}.
In Ref.~\cite{Escudero:2019gvw}, the decay of the Majoron with a keV scale mass can increase $\Delta N_{\mathrm{eff}}$ at most $0.11$ and hence help to ameliorate the Hubble tension. 

Some models with \lmlt symmetry can reproduce observed neutrino masses and mixing by introducing global U(1)$_L$ symmetry \cite{Araki:2019rmw}. 
In such models, the contributions from both the $Z'$ boson and Majoron have to be taken into account by tracking the number and energy densities of light particles in the early universe. 
In this paper, we consider solutions of the Hubble tension in a \lmlt model with a Majoron by including contributions of all light particles. 
For simplicity, we only focus on a case where the Majoron does not exist at the beginning of the universe and is created by $\nu\nu\to\phi$ after $e^\pm$ annihilation. 
In this case, $N_{\rm eff}$ can be calculated separately from the contribution of $Z'$ boson and that of $\phi$.

This paper is organized as follows. 
In section~\ref{sec:model}, we describe the \lmlt model with the global U(1)$_L$ symmetry.
In section~\ref{sec:eqs}, we derive the evolution equations of the temperature and chemical potential in the early universe. 
In section~\ref{sec:num}, we solve these equations in order to calculate the contribution of $Z^\prime$ and Majoron to $N_{\mathrm{eff}}$ and impose a constraint on $Z^\prime$ and Majoron parameter space.
Finally, we summarize our results in section~\ref{sec:sum}.

\section{\lmlt Model}
\label{sec:model}

We consider a \lmlt model which contains the global U(1)$_L$ symmetry, similarly to Ref.~\cite{Araki:2019rmw}. 
Such a model can have a keV Majoron as a pseudo Nambu-Goldstone boson (pNG boson) originated from the spontaneous symmetry breaking of the U(1)$_L$.
In addition, this model has a \lmlt gauge boson, which can explain the muon anomalous magnetic moment and the IceCube gap of cosmic neutrino flux if this gauge boson has $\mathcal{O}(10\mathchar`-100)$ MeV mass 
\cite{Araki:2014ona,Kamada:2015era, Araki:2015mya, DiFranzo:2015qea}.
As discussed in Refs.~\cite{Escudero:2019gzq,Escudero:2019gvw}, these particles can contribute to the expansion history of the early universe and have a possibility to resolve the Hubble tension.

In this section, we show the interactions between the electron, neutrino, \lmlt gauge boson $Z'$, and Majoron $\phi$, which contribute to the Hubble parameter in the early universe.

\subsection{The \lmlt Lagrangian}
\label{subsec:LmuLtau-Lagrangian}

The Lagrangian related to the \lmlt gauge boson is given by
\begin{align}
   \mathcal{L}_{Z^\prime} 
   &= -\frac{1}{4} Z^{\prime \, \rho \sigma} Z^\prime _{\rho \sigma} + \frac{1}{2} m_{Z^\prime}^2 Z^{\prime \rho} Z^\prime_\rho +g_{\mu-\tau} Z^\prime_\rho J_{\mu-\tau}^\rho \label{Lzm}~,
\end{align}
where $Z'$ denotes the \lmlt gauge boson with the field strength $Z^\prime_{\rho \sigma} = \partial_\rho Z^\prime_\sigma - \partial_\sigma Z^\prime_\rho$, and $m_{Z'}$ and $g_{\mu-\tau}$ are the \lmlt gauge boson mass and gauge coupling constant, respectively.
$J_{\mu-\tau}$ denotes the $L_\mu-L_\tau$ current and is written by
\begin{align}
 J_{\mu-\tau}^\rho&= \bar{\mu} \gamma^\rho \mu + \bar{\nu}_\mu \gamma^\rho P_L \nu_\mu - \bar{\tau} \gamma^\rho \tau - \bar{\nu}_\tau \gamma^\rho P_L \nu_\tau~.
\end{align}
At tree level, the \lmlt gauge boson interacts only with mu and tau-type leptons.

\subsection{Effective coupling with electrons}
\label{subsec:Zpee-coupling}

\begin{figure}[h]
	\begin{center}
	\includegraphics{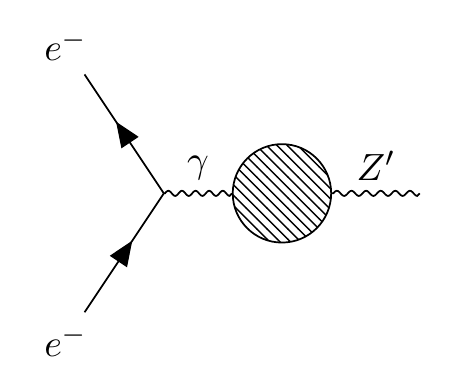}
	\caption{One-loop diagram which induces an interaction between $Z^\prime$ and electrons.\label{fig:Ze}}
	\end{center}
\end{figure}

In this model, there can be a gauge kinetic mixing $\chi$ between $Z'$ and the SM hypercharge gauge field $B$~: 
$
\mathcal{L}_{\rm mix} =-\frac{\chi}{2} B^{\rho \sigma} Z^\prime_{\rho \sigma}
$ 
where $B_{\mu \nu}$ is the field strength of $B$.
Although we assume that this kinetic mixing vanishes at some high scale for simplicity, non-zero kinetic mixing appears at one-loop level at a low energy scale.
This kinetic mixing then induces an interaction between $Z'$ and electrons through the mixing $\epsilon$ of $Z'$ with the SM photon $\gamma$ as shown in Fig~\ref{fig:Ze}, and the interaction term is described as follows~:
\begin{align}
\mathcal{L}_{Z^\prime} \supset -\epsilon e Z^\prime_\mu \bar{e} \gamma^\mu e~,
\end{align}
where $\epsilon$ is calculated by
\begin{align}
\epsilon 
\simeq \frac{e g_{\mu- \tau}}{12\pi^2} \log \frac{m_\tau^2}{m_\mu^2} 
\simeq \frac{g_{\mu-\tau}}{70}~,
\label{eq:epsilon}
\end{align}
where $e$ and $m_\ell$ are the electromagnetic charge and the mass of charged lepton $\ell$.

The partial decay widths of $Z'$ are given as follows:
\begin{align}
   \Gamma_{Z^\prime \to e^-e^+} 
   &= \frac{(\epsilon e)^2m_{Z^\prime}}{12 \pi} 
      \biggl( 1+\frac{2m_e^2}{m_{Z^\prime}^2} \biggr) \sqrt{1-\frac{4m_e^2}{m_{Z^\prime}^2}}~, \\
   \Gamma_{Z^\prime \to \nu_{\mu, \tau} \bar{\nu}_{\mu, \tau} } 
   &= \frac{g_{\mu-\tau}^2 m_{Z^\prime}}{24 \pi}~.
\end{align}
Hereafter, we assume that neutrino masses are negligible and taken to be massless.
Note that the BABAR experiment excludes the \lmlt gauge boson with $m_{Z'} > 2 m_\mu$ as a solution of the muon anomalous magnetic moment, and thus we assumed $m_{Z^\prime}< m_\mu$.

\subsection{Majoron interactions}
\label{subsec:majoron-int}

The spontaneous breaking of the global U(1)$_L$ symmetry gives rise to a Nambu-Goldstone boson, called the Majoron $\phi$.
If the global U(1)$_L$ symmetry is slightly broken, then the Majoron has a tiny mass~:
\begin{align}
\mathcal{L}_{\rm mass} = -\frac{1}{2} m_\phi^2 \phi^2~.
\end{align} 
The interaction between the Majoron and neutrinos is described by
\begin{align}
\mathcal{L}_{\rm int} =g_{\alpha \beta} \bar{\nu}_{L,\alpha} \nu_{L,\beta}^c \phi +h.c.~, 
\label{phicoupling}
\end{align}
where $g_{\alpha \beta} =g_{\beta \alpha}$ is coupling constants and $\nu_{L,\alpha}^c \equiv (\nu_{L,\alpha})^c =C \bar{\nu}_{L,\alpha}^T$ with the charge conjugation matrix $C$. 
As we will see later, this interaction can have a significant impact on the early universe.

Using the projection operator as $\nu_{L,\alpha}=P_L \nu_\alpha$, we can rewrite the Lagrangian as
\begin{align}
\mathcal{L}_{\rm int}
&= g_{\alpha \beta} \bar{\nu}_\alpha P_R C \bar{\nu}_\beta^T \phi +g_{\alpha \beta}^* \nu_\alpha^T CP_L \nu_\beta \phi \notag
\\ 
&= \sum_\alpha g_{\alpha \alpha} \bar{\nu}_\alpha P_R C \bar{\nu}_\alpha^T \phi + 2\sum_{\alpha < \beta} g_{\alpha \beta} \bar{\nu}_\alpha P_R C \bar{\nu}_\beta^T \phi +\sum_\alpha g_{\alpha \alpha}^* \nu_\alpha^T CP_L \nu_\alpha \phi +2\sum_{\alpha < \beta} g_{\alpha \beta}^* \nu_\alpha^T CP_L \nu_\beta \phi~.
\end{align}
In the first equality, we used $C(\gamma^5)^T=\gamma^5C$. 
From the above interactions, 
we obtain the decay width for $\phi \to \nu_\alpha \nu_\beta,~\phi \to \bar{\nu}_\alpha \bar{\nu}_\beta$ as
\begin{align}
\Gamma_{\phi \to \nu_\alpha \nu_\beta} = \Gamma_{\phi \to \bar{\nu}_\alpha \bar{\nu}_\beta}=\frac{|g_{\alpha \beta}|^2 m_\phi}{4\pi S_{\alpha \beta}}~.
\end{align}
Here, $S_{\alpha \beta}$ is a symmetry factor satisfying $S_{\alpha \beta} = 2 (\alpha = \beta),~S_{\alpha \beta} = 1  (\alpha \neq \beta)$.

\section{Time evolution equation of temperature and chemical potential}
\label{sec:eqs}

Here we consider the thermodynamics of the early universe in the presence of new light particles, $Z^\prime$ and the Majoron $\phi$.
In our study, we assume the following conditions~:
\begin{enumerate}
\item 
As for the parameters of $Z^\prime$, we focus on the region of $g_{\mu-\tau} \sim 10^{-4} \mathchar`- 10^{-3}$ and $m_{Z'} \sim 10$ MeV 
to solve the $(g-2)_{\mu}$ anomaly.

\item 
As for the Majoron-neutrino couplings given in Eq.~\eqref{phicoupling}, we focus on the region of $|g_{\alpha \beta}| \lesssim 10^{-7}$ in order to evade the constraints from the Big Bang Nucleosynthesis (BBN)~\cite{Escudero:2019gvw}, KamLAND-Zen~\cite{Gando:2012pj}, and SN1987A~\cite{Kachelriess:2000qc, Farzan:2002wx}.

\item 
We assume that there is no primordial abundance of Majorons, and they are produced after $e^\pm$ annihilation through the inverse decay process $\nu \nu \to \phi$\footnote{
The initial condition $n_{\phi} = 0$ 
in the early universe where $U(1)_{L}$ symmetry is restored 
would be guaranteed as follows.
Let $S$ be an original field of the Majoron when the $U(1)_{L}$ symmetry is unbroken. 
Here we consider situations where $S$ develops a vacuum expectation value (vev) 
after weak bosons decouple ($T \ll m_{Z,W}/3$).
If the field $S$ is sufficiently heavy and is not created by the decay of other fields, the number density of $S$ in the early universe is negligible. 
For example, in Ref.~\cite{Araki:2019rmw}, the field $S_{L}$ has a mass of about TeV that is greater than masses of heavy neutrinos $M_{N} \sim O(100)$ GeV and acquires a vev $\sim O(10^{-7})$ GeV. 
Thus, the initial condition $n_{\phi} = 0$ is justified. }; 
this assumption corresponds to looking at the parameter region satisfying Eq.~\eqref{eq:bou}. 
Boltzmann equations with simultaneous contributions from $Z'$ and $\phi$ are technically difficult to solve. We leave it for future work. 

\end{enumerate}
Under condition 2, the scattering and the annihilation processes of Majorons can be neglected, and only the decay and inverse decay of the Majoron $\phi \leftrightarrow \nu_\alpha \nu_\beta,~\bar{\nu}_\alpha \bar{\nu}_\beta$ are relevant to our study. 
Moreover, because of condition 1, $Z'$ becomes non-relativistic before $e^\pm$ annihilation and decays mainly into neutrinos.
On the other hand, from condition 3, Majorons are produced after $e^\pm$ annihilation.
Therefore, thermodynamics of $Z'$ and $\phi$ can be considered separately, before and after the temperature $T_\gamma \sim 10^{-2}$ MeV at which the electrons and positrons have
already annihilated.
In the following subsections, the evolution equations are derived for each period.

\subsection{Evolution equation before $e^\pm$ annihilation}

We consider the evolution equations for the universe before $e^\pm$ annihilation, at which photons, neutrinos, electrons, and $Z^\prime$ exist.
Following the previous studies \cite{Escudero:2019gzq,Escudero:2018mvt,Escudero:2020dfa}, we make the following approximations in the calculation.

\begin{enumerate}
\item[1.] All the particles follow the thermal equilibrium distribution function.

\vspace{0.3cm}
\item[2.] In the collision terms, we use the Maxwell-Boltzmann statistics.

\vspace{0.3cm}
\item[3.] Neglect the electron mass $m_e$ in the collision terms for the weak interaction processes. 

\vspace{0.3cm}
\item[4.] Neglect the chemical potentials $\mu_i$ for all the particles $i$.

\vspace{0.3cm}
\item[5.] 
The temperatures $T_i$ of a particle $i$ in the same thermal bath are equal; 
$T_\gamma = T_{e^-}$ and  
$T_{\nu_\alpha}=T_{Z^\prime} \equiv T_\nu$ for $\alpha= e,\mu,\tau$. \\

\end{enumerate}

Using these approximations, we obtain the evolution equations for the temperatures of photon $T_\gamma$ and neutrinos $T_\nu$ as follows \cite{Escudero:2019gzq}: 
\begin{align}
\frac{d T_\nu}{dt} &= 
- \biggl( \frac{\partial \rho_{\nu} }{\partial T_{\nu}} +\frac{\partial \rho_{Z^\prime}}{\partial T_\nu} \biggr)^{-1}  
\biggl[ 4H \rho_{\nu} +3H(\rho_{Z^\prime} + P_{Z^\prime})- \frac{\delta \rho_{\nu}}{\delta t}  -\frac{\delta \rho_{Z^\prime}}{\delta t} \biggr] \label{eq:TnuZ}~, \\
\frac{d T_\gamma}{dt} &= 
- \biggl( \frac{\partial \rho_\gamma}{\partial T_\gamma} + \frac{\partial \rho_e}{\partial T_\gamma} \biggr)^{-1} 
\biggl[ 4H \rho_\gamma + 3H ( \rho_e  + P_e) + \frac{\delta \rho_{\nu}}{\delta t} + \frac{\delta \rho_{Z^\prime}}{\delta t}  \biggr]~,
\label{eq:TgamZ}
\end{align}
with $\rho_i$ and $P_i$ being the energy density and pressure of particle $i$, respectively, and $H$ the Hubble parameter.
Here, the energy transfer rates in Eqs.~\eqref{eq:TnuZ} and \eqref{eq:TgamZ} are given by
\begin{align}
\frac{\delta \rho_{Z^\prime}}{\delta t} &= 
\frac{3m_{Z^\prime}^3}{2 \pi^2} \biggl[ T_\gamma K_2\biggl( \frac{m_{Z^\prime}}{T_\gamma} \biggr) -T_\nu K_2\biggl( \frac{m_{Z^\prime}}{T_\nu}\biggr) \biggr] \Gamma_{Z^\prime \to e^+e^-}~, \\
\frac{\delta \rho_{\nu}}{\delta t} &= 
\frac{4G_F^2}{\pi^5}  \Bigl[(g_{e L}^2+g_{e R}^2) +2 (g_{\mu L}^2+g_{\mu R}^2) \Bigr] F(T_\gamma,T_\nu) +\frac{2 (g_{\mu-\tau} \epsilon e)^2 }{\pi^5 m_{Z^\prime}^4}  F(T_\gamma,T_\nu)~, 
\end{align}
where $G_F$ is the Fermi coupling constant, $K_2$ is the modified Bessel function of the second kind, and $g_{e L}=1/2 + s_W^2, \; g_{e R}=s_W^2, \; g_{\mu L}=-1/2 + s_W^2$, and $g_{\mu R}=s_W^2$ with the sine of Weinberg angle $s_W \equiv \sin \theta_{W}$~.
The function $F(T_1,T_2)$ is defined as
\begin{align}
F(T_1,T_2) = 32(T_1^9-T_2^9)+56T_1^4T_2^4(T_1-T_2)~. 
\end{align}

\subsection{Evolution equation after $e^\pm$ annihilation}
\label{after-epm}

We derive the evolution equations for the universe after $e^\pm$ annihilation, at which photons, neutrinos, and the Majoron exist.
In analogy with the previous subsection, we make the following assumptions~\cite{ Escudero:2020dfa}.

\begin{enumerate}
\item[1.] All the particles follow the thermal equilibrium distribution function.

\vspace{0.3cm}
\item[2.] In the collision terms,  we use the Maxwell-Boltzmann statistics.

\vspace{0.3cm}
\item[3.] $T_{\nu_\alpha} \equiv T_\nu$ and $\mu_{\nu_\alpha} \equiv \mu_\nu  \; (\alpha = e, \mu, \tau)$. 

\end{enumerate}

Using these approximations, we obtain the evolution equations for temperature and chemical potential as follows~\footnote{
A derivation of these equations can be found in Appendix~\ref{appA}.
}~\cite{Escudero:2020dfa}~:
\begin{align}
\label{dTnu-4}
\frac{dT_\nu}{dt} &= 
\biggl(\frac{\partial n_{\nu}}{\partial \mu_{\nu}} \frac{\partial \rho_{\nu}}{\partial T_{\nu}}-\frac{\partial n_{\nu}}{\partial T_{\nu}} \frac{\partial \rho_{\nu}}{\partial \mu_{\nu}} \biggr)^{-1}  
\biggl[ -3H \biggl( (\rho_{\nu}+P_{\nu})\del{n_{\nu}}{\mu_\nu}-n_{\nu} \del{\rho_{\nu}}{\mu_\nu} \biggr) +\del{n_{\nu}}{\mu_\nu} \frac{\delta \rho_{\nu}}{\delta t} - \del{\rho_{\nu}}{\mu_\nu} \frac{\delta n_{\nu}}{\delta t} \biggr]~,  \\
\label{dmunu-4}
\frac{d\mu_\nu}{dt} &= 
-\biggl( \frac{\partial n_{\nu}}{\partial \mu_{\nu}} \frac{\partial \rho_{\nu}}{\partial T_{\nu}}-\frac{\partial n_{\nu}}{\partial T_{\nu}} \frac{\partial \rho_{\nu}}{\partial \mu_{\nu}} \biggr)^{-1} 
\biggl[ -3H \biggl( (\rho_\nu+P_\nu)\del{n_\nu}{T_\nu}-n_\nu \del{\rho_\nu}{T_\nu} \biggr) +\del{n_\nu}{T_\nu} \frac{\delta \rho_\nu}{\delta t} - \del{\rho_\nu}{T_\nu} \frac{\delta n_\nu}{\delta t} \biggr]~, \\  
\label{dTphi} 
\frac{dT_\phi}{dt} &= 
\biggl( \frac{\partial n_{\phi}}{\partial \mu_{\phi}} \frac{\partial \rho_{\phi}}{\partial T_{\phi}}-\frac{\partial n_{\phi}}{\partial T_{\phi}} \frac{\partial \rho_{\phi}}{\partial \mu_{\phi}} \biggr)^{-1}
\biggl[ -3H \biggl( (\rho_\phi+P_\phi)\del{n_\phi}{\mu_\phi}-n_\phi \del{\rho_\phi}{\mu_\phi} \biggr) +\del{n_\phi}{\mu_\phi} \frac{\delta \rho_\phi}{\delta t} - \del{\rho_\phi}{\mu_\phi} \frac{\delta n_\phi}{\delta t} \biggr]~, \\ 
\label{dmuphi} 
\frac{d\mu_\phi}{dt} &= 
-\biggl( \frac{\partial n_{\phi}}{\partial \mu_{\phi}} \frac{\partial \rho_{\phi}}{\partial T_{\phi}}-\frac{\partial n_{\phi}}{\partial T_{\phi}} \frac{\partial \rho_{\phi}}{\partial \mu_{\phi}} \biggr)^{-1} 
\biggl[ -3H \biggl( (\rho_\phi+P_\phi)\del{n_\phi}{T_\phi}-n_\phi \del{\rho_\phi}{T_\phi} \biggr)  \hspace{-2pt}+\del{n_\phi}{T_\phi} \frac{\delta \rho_\phi}{\delta t} - \hspace{-2pt} \del{\rho_\phi}{T_\phi} \frac{\delta n_\phi}{\delta t} \hspace{-1pt}\biggr]~, \\
\label{dTgam-4}
\frac{d T_\gamma}{dt} &=
- H T_\gamma~,
\end{align}
with $n_i$ being the number density of particle $i$. 
The number and energy density transfer rate of neutrinos are given by
\begin{align}
\frac{\delta n_{\nu} }{ \delta t} &= \sum_\alpha \biggl( \frac{\delta n_{\nu_\alpha }}{ \delta t }+ \frac{\delta n_{\bar{\nu}_\alpha } }{ \delta t} \biggr)~, \\
\frac{\delta \rho_{\nu} }{ \delta t} &= \sum_\alpha \biggl( \frac{\delta \rho_{\nu_\alpha }}{ \delta t }+ \frac{\delta \rho_{\bar{\nu}_\alpha } }{ \delta t} \biggr)~. 
\end{align}
Since $\phi$ and $\nu$ are no longer strongly coupled to the photon in this period, their chemical potentials are no longer guaranteed to be zero. Thus, the above evolution equations for $\mu_\nu$ and $\mu_\phi$ are indispensable.

\subsection{Calculation of the number and energy transfer rates} \label{4.4}

To solve the evolution equations for temperatures and chemical potentials, we need to calculate the number and the energy transfer rates.
For processes $\phi \leftrightarrow \nu_\alpha \nu_\beta$ and $\phi \leftrightarrow \bar{\nu}_\alpha \bar{\nu}_\beta$, the number and the energy transfer rate of $\phi$ are described as follows \cite{Escudero:2020dfa}: 
\begin{align}
\frac{\delta n_\phi}{\delta t} \bigg|_{\phi \leftrightarrow \nu_\alpha \nu_\beta}&=\frac{\delta n_\phi}{\delta t} \bigg|_{\phi \leftrightarrow \bar{\nu}_\alpha \bar{\nu}_\beta}= \frac{m_\phi^2  \Gamma_{\phi \to \nu_\alpha \nu_\beta}}{2 \pi^2} \biggl[T_\nu e^{2\mu_\nu/T_\nu} K_1\biggl(\frac{m_\phi}{T_\nu} \biggr)-T_\phi e^{\mu_\phi/T_\phi} K_1\biggl(\frac{m_\phi}{T_\phi} \biggr) \biggr] \, , 
\\ \frac{\delta \rho_\phi}{\delta t} \bigg|_{\phi \leftrightarrow \nu_\alpha \nu_\beta}&=\frac{\delta \rho_\phi}{\delta t} \bigg|_{\phi \leftrightarrow \bar{\nu}_\alpha \bar{\nu}_\beta}= \frac{m_\phi^3  \Gamma_{\phi \to \nu_\alpha \nu_\beta}}{2 \pi^2} \biggl[T_\nu e^{2\mu_\nu/T_\nu} K_2\biggl(\frac{m_\phi}{T_\nu} \biggr)-T_\phi e^{\mu_\phi/T_\phi} K_2\biggl(\frac{m_\phi}{T_\phi} \biggr) \biggr] . 
\end{align}
Actually, in addition to the decay and inverse decay of $\phi$, there also exist the scattering and the annihilation processes of  Majoron.
However, we neglect these processes because we assume $|g_{\alpha \beta}| \lesssim 10^{-7}$ as mentioned at the beginning of this section.
In this case, the number transfer rate for $\phi$ is given by
\begin{align}
\frac{\delta n_\phi}{\delta t}
&= \sum_{\alpha \leq \beta} \biggl( \frac{\delta n_\phi}{\delta t} \bigg|_{\phi \leftrightarrow \nu_\alpha \nu_\beta} +\frac{\delta n_\phi}{\delta t} \bigg|_{\phi \leftrightarrow \bar{\nu}_\alpha \bar{\nu}_\beta}\biggr) \notag \\
&= \frac{m_\phi^2  \Gamma_\phi}{2 \pi^2} \biggl[T_\nu e^{2\mu_\nu/T_\nu} K_1\biggl(\frac{m_\phi}{T_\nu} \biggr)-T_\phi e^{\mu_\phi/T_\phi} K_1\biggl(\frac{m_\phi}{T_\phi} \biggr) \biggr]~, 
\end{align}
where $\Gamma_\phi$ is the total decay width of $\phi$ given by
\begin{align}
\Gamma_\phi &\equiv \sum_{\alpha \leq \beta} (\Gamma_{\phi \to \nu_\alpha \nu_\beta}+\Gamma_{\phi \to \bar{\nu}_\alpha \bar{\nu}_\beta})
= \frac{m_\phi \lambda^2}{4 \pi}~,
\end{align}
where $\lambda^2 \equiv {\rm tr} (g^\dagger g)$.
In the same way, the energy transfer rate for $\phi$ is written as
\begin{align}
\frac{\delta \rho_\phi}{\delta t}&=\sum_{\alpha \leq \beta} \biggl( \frac{\delta \rho_\phi}{\delta t} \bigg|_{\phi \leftrightarrow \nu_\alpha \nu_\beta} +\frac{\delta \rho_\phi}{\delta t} \bigg|_{\phi \leftrightarrow \bar{\nu}_\alpha \bar{\nu}_\beta}\biggr) \notag
\\ &= \frac{m_\phi^3  \Gamma_\phi}{2 \pi^2} \biggl[T_\nu e^{2\mu_\nu/T_\nu} K_2\biggl(\frac{m_\phi}{T_\nu} \biggr)-T_\phi e^{\mu_\phi/T_\phi} K_2\biggl(\frac{m_\phi}{T_\phi} \biggr) \biggr] . 
\end{align}

The transfer rates for neutrinos, 
$\delta n_\nu/\delta t$ and $\delta \rho_\nu/\delta t$, can be obtained from the number and the energy conservation law.
In the present case, the physics does not depend on a basis of neutrinos, because the neutrino masses are neglected. 
This is understood from the fact that $\Gamma_\phi$ depends on $g_{\alpha \beta}$ only in the form $\mathrm{tr}(g^\dag g)$.
Thus, without loss of generality, we can assume that $g_{\alpha \beta}$ has only diagonal components, and the number conservation is expressed as
\begin{align}
\frac{\delta n_{\nu_\alpha}}{\delta t} \bigg|_{\phi \leftrightarrow \nu_\alpha \nu_\alpha}=-2\frac{\delta n_\phi}{\delta t} \bigg|_{\phi \leftrightarrow \nu_\alpha \nu_\alpha}~,
\end{align}
which leads to
\begin{align}
\frac{\delta n_{\nu}}{\delta t} & = \sum_\alpha \biggl( \frac{\delta n_{\nu_\alpha}}{\delta t} \bigg|_{\phi \leftrightarrow \nu_\alpha \nu_\alpha}+\frac{\delta n_{\bar{\nu}_\alpha}}{\delta t} \bigg|_{\phi \leftrightarrow \bar{\nu}_\alpha \bar{\nu}_\alpha}\biggr) 
=-2 \frac{\delta n_\phi}{\delta t}~. 
\end{align}
On the other hand, the energy conservation leads to %
\begin{align}
\frac{\delta \rho_{\nu_\alpha}}{\delta t} \bigg|_{\phi \leftrightarrow \nu_\alpha \nu_\alpha}=-\frac{\delta \rho_\phi}{\delta t} \bigg|_{\phi \leftrightarrow \nu_\alpha \nu_\alpha}~.
\label{eq:36}
\end{align}
From Eq.~\eqref{eq:36}, $\delta \rho_\nu/ \delta t$ is found to be
\begin{align}
\frac{\delta \rho_{\nu}}{\delta t} & = \sum_\alpha \biggl( \frac{\delta \rho_{\nu_\alpha}}{\delta t} \bigg|_{\phi \leftrightarrow \nu_\alpha \nu_\alpha}+\frac{\delta \rho_{\bar{\nu}_\alpha}}{\delta t} \bigg|_{\phi \leftrightarrow \bar{\nu}_\alpha \bar{\nu}_\alpha}\biggr) 
=-\frac{\delta \rho_\phi}{\delta t}~.
\end{align}

\section{Numerical calculation}
\label{sec:num}

In this section, we discuss the initial conditions and the parameters for the evolution equations of temperatures and the chemical potentials derived in the previous section and show the numerical results.
The codes for calculations are partially based on {\tt NUDEC\_BSM}~\cite{Escudero:2020dfa}.

\subsection{Initial conditions and integration range} 
\label{4.6.1}

\subsubsection*{Before $e^{\pm}$ annihilation}

We solve the system of differential equations \eqref{eq:TnuZ} and \eqref{eq:TgamZ} starting from $T_\gamma=T_\nu=20$ MeV at which all the particles are in thermal equilibrium, to $T_\gamma \sim 10^{-2}$ MeV where the $e^\pm$ annihilation has taken place.

\subsubsection*{After $e^{\pm}$ annihilation}

Let us consider solving the system of differential equations (\ref{dTnu-4})-(\ref{dTgam-4}) from the temperature where the Majoron hardly exists.
To see when the Majoron can be produced in the early universe, we can consider $\langle \Gamma_{\nu \nu \to \phi} \rangle/H$,
where $\langle \Gamma_{\nu \nu \to \phi} \rangle$ is the thermally averaged neutrino inverse decay rate, and $H$ is the Hubble rate.
The ratio $\langle \Gamma_{\nu \nu \to \phi} \rangle/H$ is written as \cite{Escudero:2020dfa}
\begin{align}
\label{eq:Gam}
\frac{\langle \Gamma_{\nu \nu \to \phi} \rangle}{H} &= \frac{1}{81K_1(3)} \Gamma_{\mathrm{eff}} \biggl( \frac{m_\phi}{T_\nu} \biggr)^4 K_1\biggl( \frac{m_\phi}{T_\nu} \biggr)~, \\
\label{eq:Geff}
\Gamma_{\mathrm{eff}} &\equiv \frac{\langle \Gamma_{\nu \nu \to \phi} \rangle}{H} \bigg|_{T_\nu=m_\phi/3}\simeq \biggl( \frac{\lambda}{4 \times 10^{-12}} \biggr)^2 \biggl( \frac{\mathrm{keV}}{m_\phi} \biggr)~.
\end{align} 
This is illustrated in Figure 2 in \cite{Escudero:2020dfa}.
Imposing $\langle \Gamma_{\nu \nu \to \phi} \rangle/H < 10^{-4}$, we obtain the condition for $T_{\nu}$ as follows~:
\begin{align}
\frac{T_\nu}{m_\phi} > \biggl( \frac{\Gamma_{\mathrm{eff}}}{81K_1(3) \times 10^{-4}} \biggr)^{1/3} \simeq 10 \, \Gamma_{\mathrm{eff}}^{\; 1/3}~.
\end{align}
Here, we use the approximation $K_1(x) \sim 1/x \; (\mathrm{for} \; x<1)$ because the situation with $T_\nu/m_\phi >1$ is what we want to consider. 
If we set the range $\Gamma_{\mathrm{eff}}\leq 10^3$, the initial value of $T_\nu$ should satisfy $T_\nu \gtrsim 100m_\phi$.
Thus, we will take 
\begin{align}
T_\nu =100 \, m_\phi \, , 
\end{align}
as the initial condition for $T_\nu$. 
As the initial condition for $T_\gamma/T_\nu$, we use the numerical values after $e^\pm$ annihilation ($T_\gamma \simeq 10^{-2}$ MeV) obtained by solving equations \eqref{eq:TnuZ} and \eqref{eq:TgamZ}.

The remaining initial conditions are determined so that $\rho_\phi/\rho_\nu < 10^{-12}$ is satisfied.
Since the Majoron is  ultra-relativistic in $T_\nu=100 \, m_\phi$, we can treat the Majoron as a massless particle, so $\rho_\phi/\rho_\nu$ is written by
\begin{align}
\frac{\rho_\phi}{\rho_\nu} = \frac{1}{6} \biggl( \frac{T_\phi}{T_\nu} \biggr)^4 \frac{\mathrm{Li}_4(e^{\mu_\phi/T_\phi})}{-\mathrm{Li}_4(-e^{\mu_\nu/T_\nu})} 
= \frac{4}{21} \biggl( \frac{T_\phi}{T_\nu} \biggr)^4 \biggl(1+a \frac{\mu_\phi}{T_\phi}- \frac{6}{7} a \frac{\mu_\nu}{T_\nu} +\cdots  \biggr) . 
\end{align}
Here, $\mathrm{Li}_s(z)$ is Polylogarithm and $a \equiv \zeta(3)/\zeta(4) \sim 1.2/1.08 \sim 1.1$. 
Therefore, to satisfy $\rho_\phi/\rho_\nu <10^{-12}$, 
the parameters should be
\begin{align}
\frac{T_\phi}{T_\nu} \lesssim 10^{-3}, \quad \bigg| \frac{\mu_\phi}{T_\phi} \bigg| < 1 , \quad \bigg| \frac{\mu_\nu}{T_\nu} \bigg| < 1  \, ,
\label{bou1}
\end{align}
This means that the condition for $\mu_\phi$ is
\begin{align}
\bigg| \frac{\mu_\phi}{T_\nu} \bigg| =\bigg| \frac{\mu_\phi}{T_\phi} \bigg|  \; \frac{T_\phi}{T_\nu} <  \frac{T_\phi}{T_\nu} \lesssim 10^{-3} \, . 
\end{align}
Furthermore, since the Majoron is a boson, $\mu_\phi$ must satisfy %
\begin{align}
\frac{\mu_\phi}{T_\nu} <  \frac{m_\phi}{T_\nu}=10^{-2} , 
\label{bou2}
\end{align}
from $\mu_\phi \leq m_\phi$.
Here, the equality sign is removed because the Bose-Einstein condensation cannot occur due to the very small number density of Majoron.

As the initial conditions that satisfies Eqs.~(\ref{bou1})-(\ref{bou2}), 
in this paper we take them as 
\begin{align}
\frac{T_\phi}{T_\nu} = 10^{-3}, ~~~ \; \frac{\mu_\nu}{T_\nu}=-10^{-4}, ~~~ \; \frac{\mu_\phi}{T_\nu} =-10^{-5} , 
\end{align}
according to \cite{Escudero:2020dfa}.
The differential equations are solved until $\rho_\phi/\rho_\nu <10^{-6}$, when the Majoron has completely decayed away.~\footnote{
For $\Gamma_{\mathrm{eff}}<0.1$, we solve the equations until $\rho_\phi/\rho_\nu <10^{-7}$ 
because it takes a long time for the Majoron to decay.
}

\subsection{Parameters}

As mentioned before, we consider the case where the Majoron does not exist in the very early universe and is created after  $e^\pm$ annihilation $(T_\gamma \lesssim 10^{-2} \mathrm{MeV})$.
To realize this situation, the parameters of the Majoron must satisfy the following conditions~:
\begin{itemize}
\item The Majoron production is most active after $e^\pm$ annihilation.

\item Shortly after $e^\pm$ have annihilated $(T_\gamma \simeq 10^{-2} \mathrm{MeV})$, the Majoron production is not yet effective.
\end{itemize}

Since $\langle \Gamma_{\nu \nu \to \phi}\rangle/H$ is maximal when $T_\nu \simeq m_\phi/3$ \cite{Escudero:2020dfa}, the above conditions are expressed as
\begin{align}
m_\phi /3 <10^{-2}~\mathrm{MeV} ,  ~~~ 
\frac{\langle \Gamma_{\nu \nu \to \phi} \rangle}{H} \bigg|_{T_\nu=10^{-2}~\mathrm{MeV}} < 1~. 
 \label{eq:bou}
\end{align}

\subsection{Results}

Here, we show the results of solving the evolution equations derived in the previous section.
In this study, the deviation of $N_{\mathrm{eff}}$ from the standard value occurs two times, namely before and after the $e^\pm$ annihilation.
Thus, it is convenient to write $N_{\mathrm{eff}}$ as
\begin{align}
N_{\mathrm{eff}} 
= N_{\mathrm{eff}}^\prime
+ \Delta N_{\mathrm{eff}}^\prime \, . \label{Neff}
\end{align}
Here, $N_{\mathrm{eff}}^\prime$ describes the effective number of neutrino species determined at $T_\nu/T_\gamma=const.$ soon after $e^\pm$ annihilation and is defined as 
\begin{align}
N_{\mathrm{eff}}^\prime &= 3\left(\frac{11}{4}\right)^{4 / 3}\left(\frac{T_{\nu}}{T_{\gamma}}\right)^{4} \bigg|_{T_\gamma \simeq 10^{-2} \mathrm{MeV}}  \, .
\end{align}
On the other hand, $\Delta N_{\mathrm{eff}}^\prime$ represents the change in the effective number of neutrino species due to the Majoron production after  $e^\pm$ annihilation.
As we will see later, $N_{\mathrm{eff}}^\prime$ and $\Delta N_{\mathrm{eff}}^\prime$ are not completely independent, and $\Delta N_{\mathrm{eff}}^\prime$ slightly depends on $N_{\mathrm{eff }}^\prime$.

\begin{figure}[ht]
\vspace{-36pt}
	\begin{center}
	\includegraphics{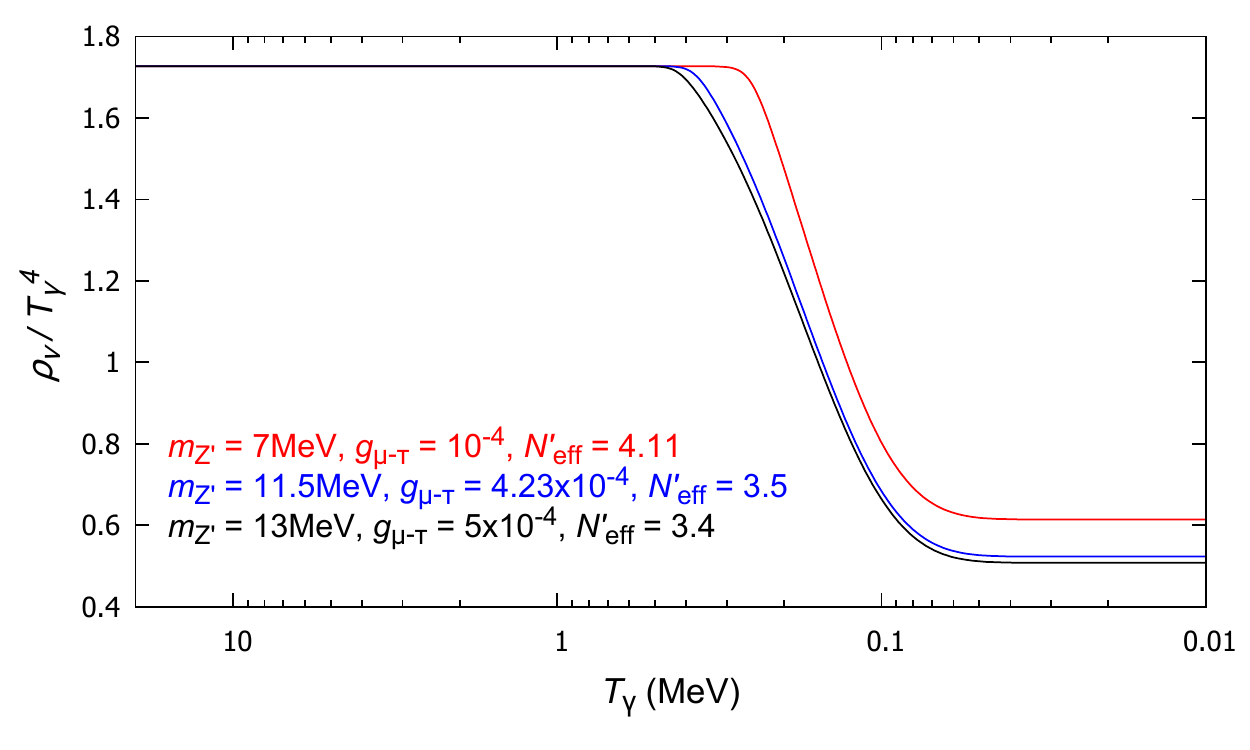}
    \caption{ 
	The evolution of the neutrino energy density for some $Z^\prime$ parameter.\label{fig.Z}}
	\end{center}
\end{figure}

Figure \ref{fig.Z} shows the evolution of the neutrino temperature obtained by solving Eqs.~\eqref{eq:TnuZ} and \eqref{eq:TgamZ}.
As can be seen from this figure, the value of $N_{\mathrm{eff}}$ is slightly larger than that of the SM $N_{\mathrm{eff}}^{\mathrm{SM}} \simeq 3.045$ \cite{deSalas:2016ztq,Akita:2020szl} due to the new gauge boson $Z^\prime$.

\begin{figure}[h]
	\begin{center}
	\includegraphics{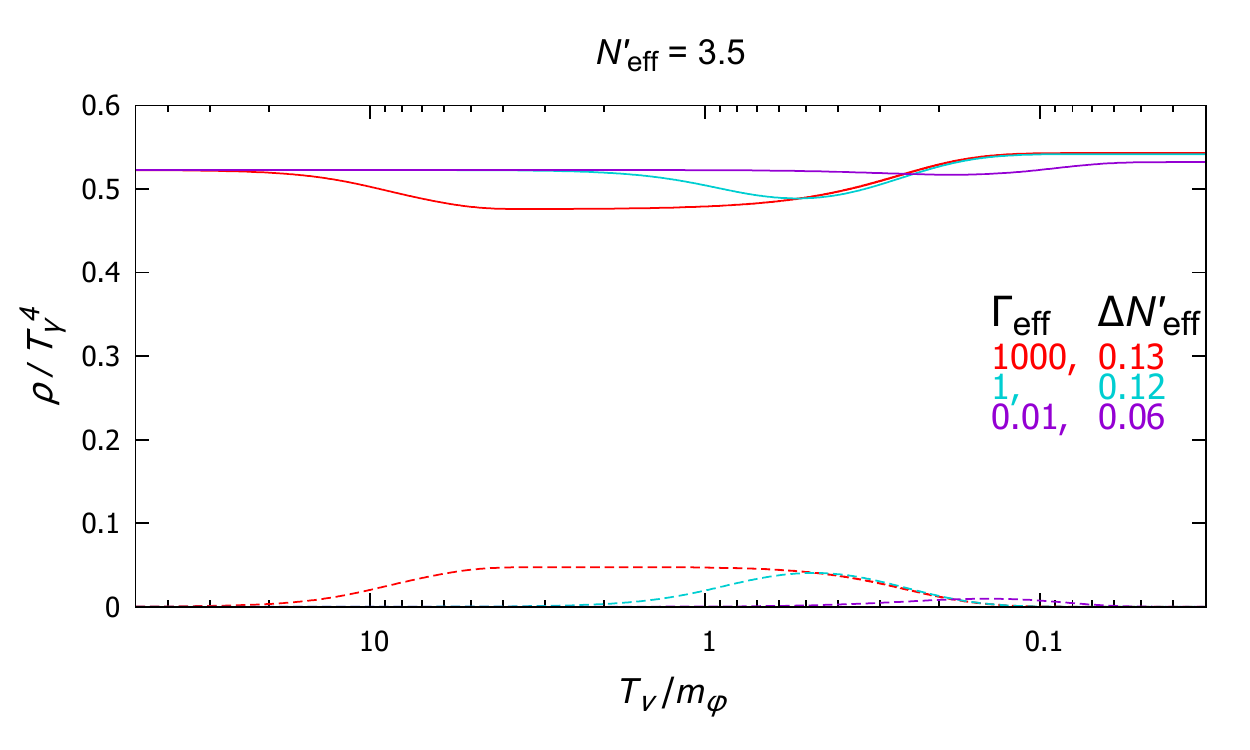}
	\caption{
	The evolution of neutrino (solid line) and Majoron (dashed line) energy density for the case of $N_{\mathrm{eff}}^\prime =3.5$. \\
	\label{nM}}
	\vspace{20pt}
	\end{center}
	\begin{center}
	\includegraphics{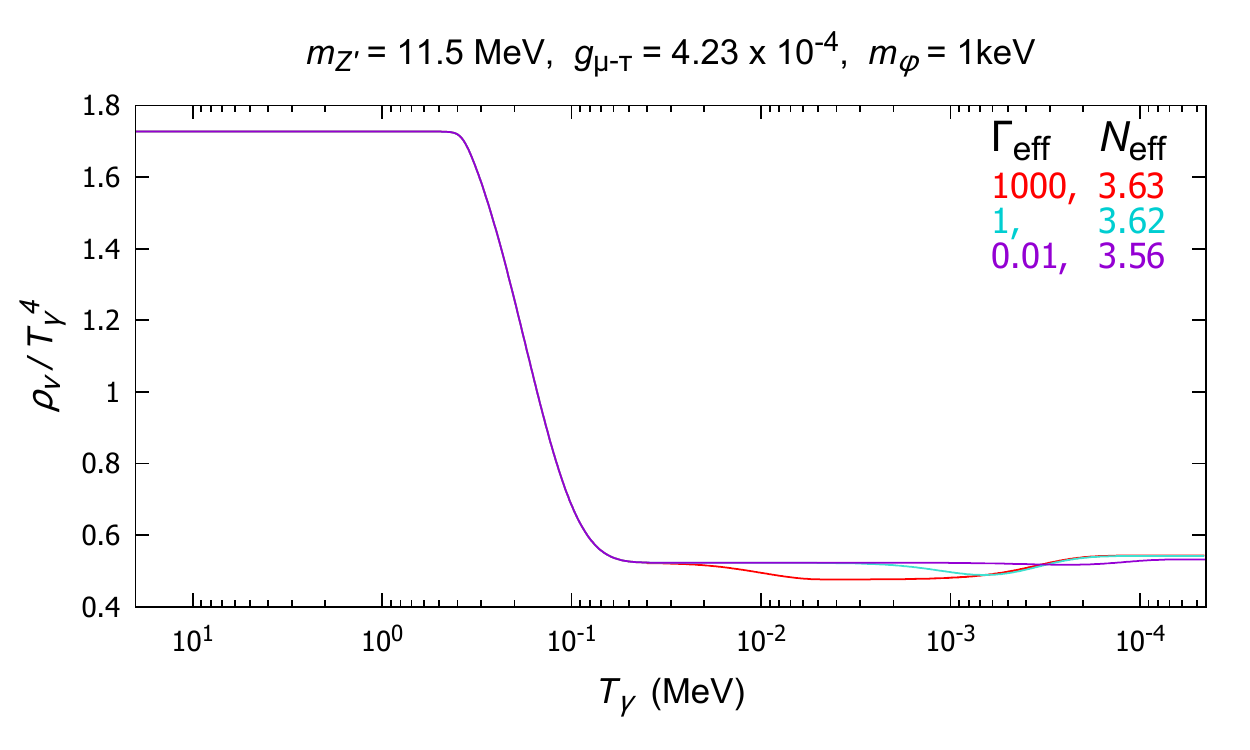}
	\caption{
	The evolution of the neutrino energy density for the case of $N_{\mathrm{eff}}^\prime=3.5, \; m_\phi=1$ keV.
	\label{fig4.3}}
	\end{center}
\end{figure}

Figure \ref{nM} shows the evolution of the neutrino energy density and the Majoron energy density for the case of $N_{\mathrm{eff}}^\prime =3.5$.
This figure shows that for $\Gamma_{\mathrm{eff}}\gtrsim 1$, the Majoron begins to be produced by $\nu\nu\to\phi$ when the temperature reaches $T_\nu\gtrsim m_\phi$.
After that, neutrinos and Majoron gradually reach the thermal equilibrium.
This corresponds to the gently sloping area around the peak in the Figure \ref{nM}.
Since the net energy transfer due to $\phi\leftrightarrow \nu\nu$ is negligibly small, the evolution of the energy densities can be determined by the following Boltzmann equations:
\begin{align}
\frac{d \rho_\nu}{dt} &+4H \rho_\nu =0 \, , 
\\ \frac{d \rho_\phi}{dt} &+3H (\rho_\phi+P_\phi) =0 \, .
\end{align}
At temperature $T_\nu \lesssim m_\phi$, the Majoron becomes non-relativistic and $\rho_\phi$ becomes much larger than $P_\phi$. 
Consequently, the energy densities are derived as follows~:
\begin{align}
\rho_\nu &\propto R^{-4} , ~~~ \rho_\phi \propto R^{-3} , 
\end{align}
where $R$ is the scale factor.
Therefore, the difference between $\rho_\nu$ and $\rho_\phi$ occurs as the universe expands.
At temperature $T_\nu \simeq m_\phi/3$, Majorons start to decay into neutrinos.
Since the neutrinos produced by this decay are more energetic than the existing neutrinos, the overall neutrino energy density slightly increases, resulting in a slightly larger $N_{\mathrm{eff}}$.

\clearpage

Figure \ref{fig4.3} shows the evolution of the neutrino energy density for the case of $N_{\mathrm{eff}}^\prime=3.5, \; m_\phi=1$ keV.
This figure is obtained by connecting Figure \ref{fig.Z} and Figure \ref{nM} at $T_\gamma \simeq 10^{-2}$ MeV smoothly.

\begin{figure}[hb]
	\begin{center}
	\includegraphics{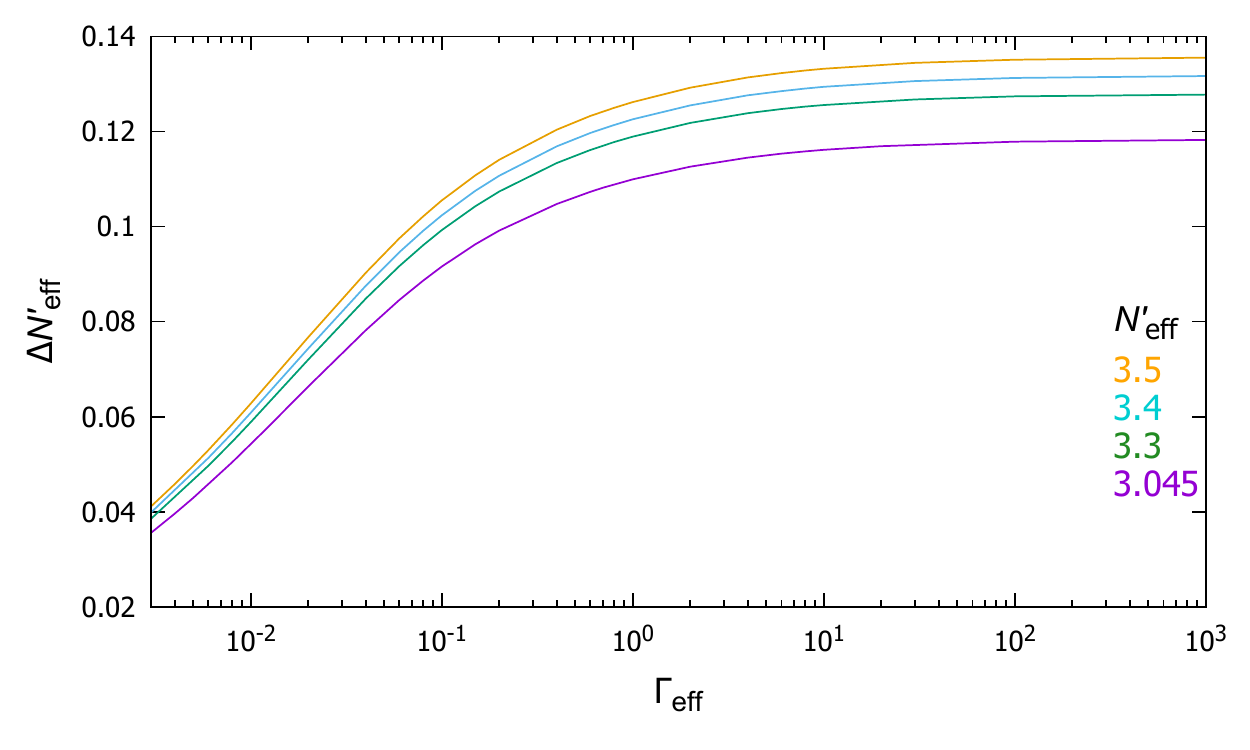}
	\caption{
	The $\Gamma_{\mathrm{eff}}$ dependence of $\Delta N_{\mathrm{eff}}^\prime$ for some $N_{\mathrm{eff}}^\prime$.
	\label{fig4.4}}
	\end{center}
\end{figure}

Figure \ref{fig4.4} shows the $\Gamma_{\mathrm{eff}}$ dependence of $\Delta N_{\mathrm{eff}}^\prime$ for some $N_{\mathrm{eff}}^\prime$. 
The parameters $\Delta N_{\rm eff}^\prime$ and $N_{\rm eff}^\prime$ are not completely independent, and
$\Delta N_{\rm eff}^\prime$ slightly depends on $N_{\rm eff}^\prime$.
As you can see, $\Delta N_{\mathrm{eff}}^\prime$ becomes larger for larger $N_{\mathrm{eff}}^\prime$. 
The reason is as follows~:
A large $N_{\mathrm{eff}}^\prime$ corresponds to a large number of neutrinos after $e^\pm$ annihilation.
For $\Gamma_{\mathrm{eff}}\gtrsim 1$, which corresponds to the case where the thermal equilibrium between the Majoron and neutrino is reached due to $\phi \leftrightarrow \nu \nu$,
this process acts to equalize the number of neutrinos and Majorons. 
Thus, for the larger number of neutrinos after $e^\pm$ annihilation, the more neutrinos are converted to the Majorons.
As a result, the neutrino energy density at $T_\nu \ll m_\phi$ becomes larger, yielding an increase in $\Delta N_{\mathrm{eff}}^\prime$.
On the other hand, for $\Gamma_{\mathrm{eff}}\ll1$, the thermal equilibrium is not achieved between $\nu$ and $\phi$, but a small amount of Majoron is produced by $\nu \nu \to \phi$.
This process occurs more often for a larger number of neutrinos after $e^\pm$ annihilation. 
Thus, the production of Majoron increases slightly and it leads to an increase in $\Delta N_{\mathrm{eff}}^\prime$. 
Note that  the contribution of Majoron $\Delta N_{\mathrm{eff}}^\prime$ cannot be larger than $\simeq 0.12$ in the case of the SM $N_{\mathrm{eff}}^\prime = 3.045$. 

\vspace{36pt}

\clearpage

\begin{figure}[h]
	\begin{center}
	\includegraphics[scale=0.4]{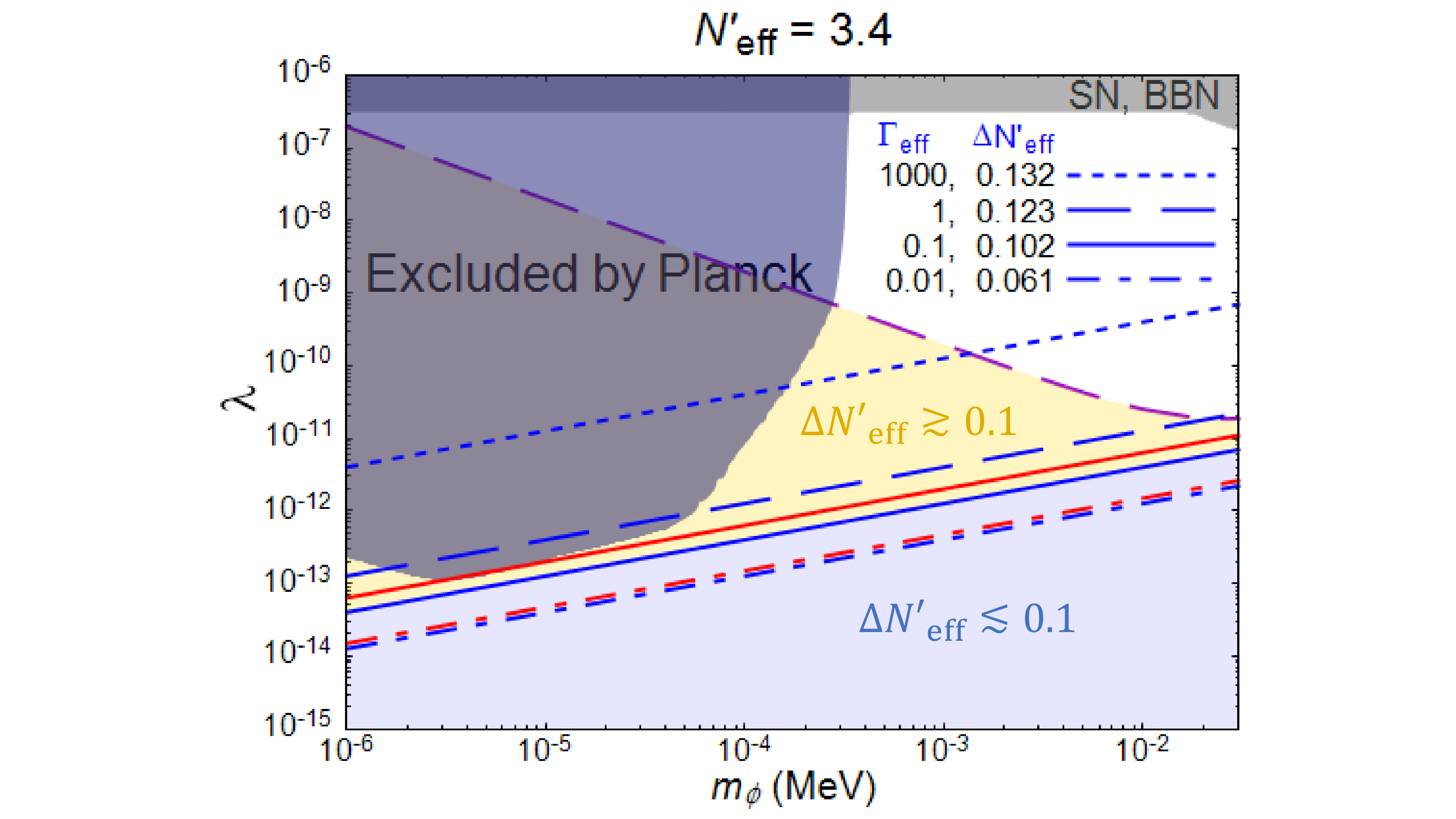}
	\caption{
	Parameter space of the Majoron in the presence of $Z^\prime$ that realizes $N_{\mathrm{eff}}^\prime=3.4$.
	The solid and dotted blue lines are the contour lines of $\Delta N_{\mathrm{eff}}^\prime \; (\Gamma_{\mathrm{eff}})$.
        The solid and dotted red lines represent the same contour lines without $Z'$ boson ($N_{\rm eff}' = 3.045$). 
	The area below the dashed purple line corresponds to one which satisfies Eq.~\eqref{eq:bou}.
	The gold region represents the region where $\Delta N_{\mathrm{eff}}^\prime \gtrsim 0.1$ holds.
	The blue region represents the parameter region where Hubble tensions can be resolved ($ 3.4 \lesssim N_{\mathrm{eff}} \lesssim 3.5$). The dark blue region is excluded by Planck 2018 data \cite{Escudero:2019gvw}.
	The gray region is excluded by SN1987A \cite{Kachelriess:2000qc,Farzan:2002wx}, Big Bang Nucleosynthesis (BBN) \cite{Escudero:2019gvw}.
     The white region cannot be treated in this paper. 
     \label{fig4.5}}
	\includegraphics{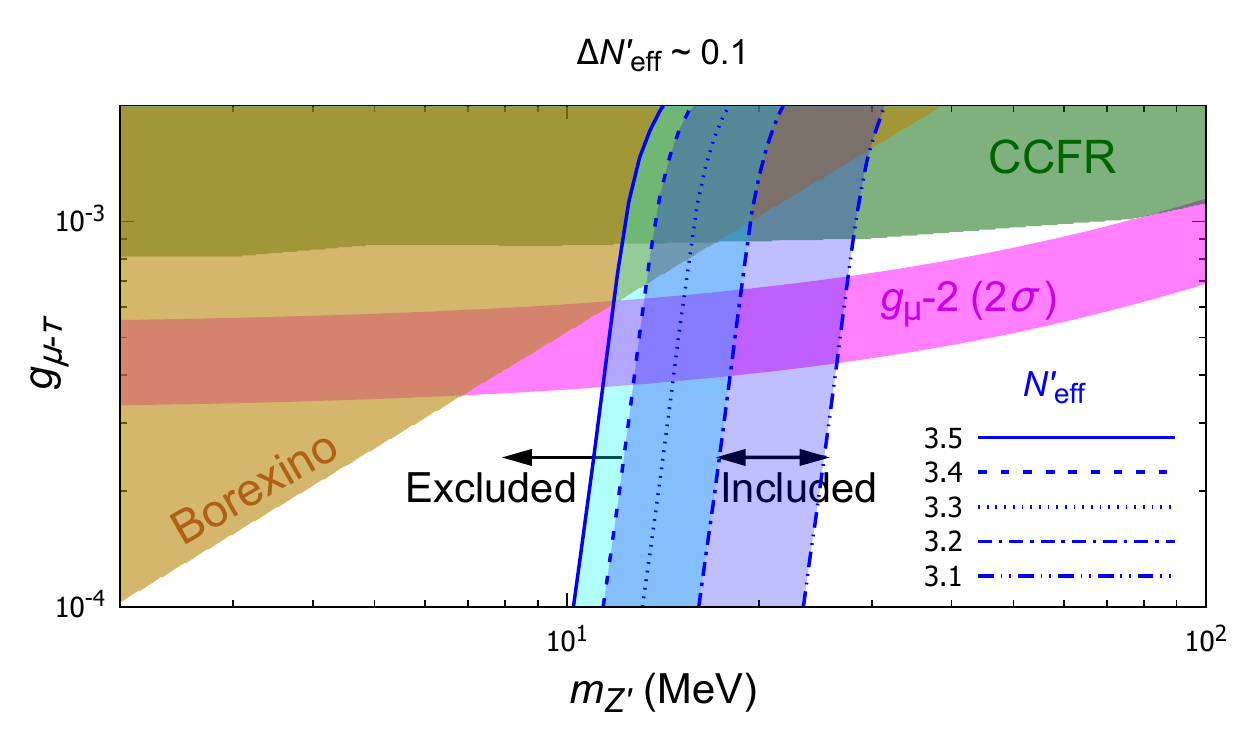}
	\caption{
	The $Z^\prime$ parameter space near the region where the $(g-2)_{\mu}$ anomaly can be resolved.
  The region between the solid and dashed dotted line (${3.2  \lesssim N'_{\mathrm{eff}} \lesssim 3.5}$) represents the region where the Hubble tension can be resolved only by $Z'$ boson. 
 The region between the dashed and dashed double-dotted line (${3.1  \lesssim N'_{\mathrm{eff}} \lesssim 3.4}$) represents the same region  in the presence of Majoron that realizes $\Delta N_{\mathrm{eff}}^\prime \simeq 0.1$.
	The magenta band represents the region where the $(g-2)_{\mu}$ anomaly can be resolved within $2\sigma$ level \cite{Muong-2:2021ojo}.
	The brown and green regions are excluded by the Borexino and CCFR experiments, respectively \cite{Araki:2017wyg}.
	\label{fig4.6}}
	\end{center}
\end{figure}

\clearpage

Using $N_{\mathrm{eff}}^\prime$ and $\Delta N_{\mathrm{eff}}^\prime$ defined above, we can write $N_{\mathrm{eff}}$ as Eq.~(\ref{Neff}).
If we fix either $N_{\mathrm{eff}}^\prime$ or $\Delta N_{\mathrm{eff}}^\prime$, a constraint can be imposed on the other parameter by using the constraint from Planck 2018~: $N_{\mathrm{eff}}=3.27\pm 0.15$ with 68\% C.L.\cite{Aghanim:2018eyx}.
Although the various patterns are possible, we will only discuss the following two cases.

\begin{enumerate}
\item For $N_{\mathrm{eff}}^\prime \simeq 3.4$.

Figure \ref{fig4.5} shows the parameter space of the Majoron in the presence of $Z^\prime$ that realizes  $N_{\mathrm{eff}}^\prime=3.4$.
The solid and dotted blue lines are the contour lines of $\Delta N_{\mathrm{eff}}^\prime \; (\Gamma_{\mathrm{eff}})$.
        The solid and dotted red lines represent the same contour lines without $Z'$ boson ($N_{\rm eff}' = 3.045$).  
The area below the dashed  purple line corresponds to Eq.~(\ref{eq:bou}).
The blue region ($\Delta N_{\mathrm{eff}}^\prime \lesssim 0.1$) represents the region where the Hubble tension can be resolved ($ 3.4 \lesssim N_{\mathrm{eff}} \lesssim 3.5$).
The lower limit of the mass of the Majoron is taken to be $10^{-6}$ MeV because neutrino masses are not negligible below this value. 
The upper limit of Majoron mass ($3 \times 10^{-2}$ MeV) corresponds to the first condition in Eq.~(\ref{eq:bou}) $m_\phi /3 <10^{-2}$ MeV. 
If $Z^\prime$ boson is in the parameter region where $(g-2)_{\mu}$ anomaly can be solved,  the Hubble tension and the $(g-2)_{\mu}$ anomaly can be resolved simultaneously in the blue region.
Furthermore, the gold region above the contour line of $\Delta N_{\mathrm{eff}}^\prime = 0.1$ is excluded at more than $2\sigma$ level.

\item For $\Delta N_{\mathrm{eff}}^\prime \simeq 0.1$.

Figure \ref{fig4.6} shows the $Z^\prime$ parameter space near the region where the $(g-2)_{\mu}$ anomaly can be resolved.
 The region between the solid and dashed dotted line (${3.2  \lesssim N'_{\mathrm{eff}} \lesssim 3.5}$) represents the region where the Hubble tension can be resolved only by $Z'$ boson, as in previous studies ({\it e.g.}, Fig. 5 in \cite{Escudero:2019gzq}). 
 The region between the dashed and dashed double-dotted line (${3.1  \lesssim N'_{\mathrm{eff}} \lesssim 3.4}$) represents the same region  in the presence of Majoron that realizes $\Delta N_{\mathrm{eff}}^\prime \simeq 0.1$. 
 In this case, the parameter region where the Hubble tension can be resolved is slightly shifted toward the larger value of $m_{Z^\prime}$. 
   As a result, a new allowed region emerges for larger $m_{Z'}$.
A choice of parameters $m_{Z^\prime}\simeq 13-26 \, \mathrm{MeV}$ and  $g_{\mu-\tau}\simeq (3.6-7)\times 10^{-4}$ can resolve the Hubble tension and $(g-2)_{\mu}$ anomaly simultaneously  in the presence of Majoron. 
The region to the left of the $N_{\mathrm{eff}}^\prime = 3.4$ contour line is excluded at more than $2\sigma$ level.  

\end{enumerate}

\section{Summary}
\label{sec:sum}


In this paper, we explored possibilities of resolving the Hubble tension and $(g-2)_{\mu}$ anomaly simultaneously in realistic \lmlt models that can explain the origin of neutrino mass.
In these models, there is a new light gauge boson $Z^\prime$ and a new light scalar, the Majoron $\phi$. It arises from the spontaneous breaking of the global $U(1)_L$ symmetry and weakly couples to neutrinos. 
The parameters of $Z^\prime$ boson are set to  be $10^{-3} \gtrsim g_{\mu-\tau}\gtrsim 10^{-4}, \; m_{Z^\prime}\simeq 10 \, $MeV, neighborhoods of  region that can resolve the $(g-2)_{\mu}$ anomaly.

We only focused on a case where the Majoron does not exist at the beginning of the universe and is created by $\nu\nu\to \phi$ after $e^\pm$  annihilation.
In this case, contributions of $Z'$ and $\phi$ to the effective number $N_{\rm eff}$ can be calculated independently.
Thus, it is convenient to write $N_{\rm eff}$ as $N_{\mathrm{eff}}=N_{\mathrm{eff}}^\prime+\Delta N_{\mathrm{eff}}^\prime$,  a sum of the effective number after $e^\pm$ annihilations $N_{\mathrm{eff}}^\prime$ and its change due to the Majoron $\Delta N_{\mathrm{eff }}^\prime$. 
The effective number $N_{\mathrm{eff}}$ is evaluated by evolution equations of temperatures and the chemical potentials of light particles in each period.

For simplicity, the following two cases are discussed.
First,  we explored the parameter space of the Majoron in the presence of $Z^\prime$ that realizes  $N_{\mathrm{eff}}^\prime=3.4$.
In this case, the Hubble tension can be resolved ($N_{\mathrm{eff}}\simeq 3.4-3.5$) in the wide region of the parameter space where $\Delta N_{\mathrm{eff}}^\prime \lesssim 0.1 \; (\lambda \lesssim 10^{-12}-10^{-14})$ holds. 
On the other hand, the region with $\Delta N_{\mathrm{eff}}^\prime \gtrsim 0.1$ is excluded at more than $2\sigma$ level.
In the second case, we surveyed the parameter region of $Z'$ where the Hubble tension can be resolved in the presence of Majoron that realizes $\Delta N_{\mathrm{eff}}^\prime \simeq 0.1$. 
A choice of parameters $m_{Z^\prime}\simeq 13-26 \, \mathrm{MeV} , \; g_{\mu-\tau}\simeq (3.6-7)\times 10^{-4}$ that corresponds to $N_{\mathrm{eff}}^\prime \simeq 3.1-3.4$ can resolve the Hubble tension and $(g-2)_{\mu}$ anomaly simultaneously. 
On the other hand, the region with $m_{Z^\prime} \lesssim 10$MeV is excluded at more than $2\sigma$ level.

As a result, 
we found that the heavier $m_{Z^\prime}$ results in the smaller $N_{\mathrm{eff}}^\prime$ and  requires the larger $\Delta N_{\mathrm{eff}}^\prime$ to resolve the Hubbel tension. 
Therefore, compared to previous studies, the parameter region where the Hubble tension can be resolved is slightly shifted toward the larger value of $m_{Z^\prime}$.
Note that $N_{\mathrm{eff}}^\prime$ and $\Delta N_{\mathrm{eff}}^\prime$ are not completely independent, and $\Delta N_{\mathrm{eff}}^\prime$ slightly depends on $N_{\mathrm{eff }}^\prime$.

Finally, Boltzmann equations with simultaneous contributions from $Z'$ and $\phi$ are more difficult to solve. We leave it for future work. 

\vspace{12pt}

\section*{Acknowledgments}
This work was supported by JSPS KAKENHI Grants 
No.~JP18H01210 (T.A., J.S., T.S., M.J.S.Y), 
No.~JP19J13812 (K.A.), 
No.~JP18K03651 (T.S.), 
No.~20K14459, (M.J.S.Y),  
and MEXT KAKENHI Grant No.~JP18H05543 (J.S., T.S., M.J.S.Y).


\appendix

\section{Derivation of the evolution equation after $e^\pm$ annihilation}
\label{appA}

Here, we derive the evolution equations (\ref{dTnu-4}) to (\ref{dmuphi}) after $e^\pm$ annihilation.
First of all, the evolution equations for the temperature $T_a$ and chemical potential $\mu_a$ of a particle species $a$ that follows the thermal equilibrium distribution function are given by \cite{Escudero:2020dfa}
\begin{align}
\frac{dT_a}{dt}&= \biggl(\frac{\partial n_{a}}{\partial \mu_{a}} \frac{\partial \rho_{a}}{\partial T_{a}}-\frac{\partial n_{a}}{\partial T_{a}} \frac{\partial \rho_{a}}{\partial \mu_{a}} \biggr)^{-1} \biggl[ -3H \biggl( (\rho_a+P_a)\del{n_a}{\mu_a}-n_a \del{\rho_a}{\mu_a} \biggr) +\del{n_a}{\mu_a} \frac{\delta \rho_a}{\delta t} - \del{\rho_a}{\mu_a} \frac{\delta n_a}{\delta t} \biggr]~, \label{dTa}
\\ \frac{d\mu_a}{dt}&= -\biggl( \frac{\partial n_{a}}{\partial \mu_{a}} \frac{\partial \rho_{a}}{\partial T_{a}}-\frac{\partial n_{a}}{\partial T_{a}} \frac{\partial \rho_{a}}{\partial \mu_{a}} \biggr)^{-1}  \biggl[ -3H \biggl( (\rho_a+P_a)\del{n_a}{T_a}-n_a \del{\rho_a}{T_a} \biggr) +\del{n_a}{T_a} \frac{\delta \rho_a}{\delta t} - \del{\rho_a}{T_a} \frac{\delta n_a}{\delta t} \biggr] .
\label{dmua}
\end{align}
In Eqs.~\eqref{dTa} and \eqref{dmua}, $n_a, \; \rho_a, \; P_a$ are the particle number density, energy density, and pressure of a particle species $a$, respectively.
From the approximations 3 in subsection \ref{after-epm} and $T_{\nu_\alpha}=T_{\bar{\nu}_\alpha}, \; \mu_{\nu_\alpha}=\mu_{\bar{\nu}_\alpha}$, 
Eq.~(\ref{dTa}) for the neutrino and antineutrino leads to 
\begin{align}
 \frac{dT_\nu}{dt}&= \biggl( \frac{\partial n_{\nu_\alpha}}{\partial \mu_{\nu}} \frac{\partial \rho_{\nu_\alpha}}{\partial T_{\nu}}-\frac{\partial n_{\nu_\alpha}}{\partial T_{\nu}} \frac{\partial \rho_{\nu_\alpha}}{\partial \mu_{\nu}} \biggr)^{-1} \biggl[ -3H \biggl( (\rho_{\nu_\alpha}+P_{\nu_\alpha})\del{n_{\nu_\alpha}}{\mu_\nu}-n_{\nu_\alpha} \del{\rho_{\nu_\alpha}}{\mu_\nu} \biggr) +\del{n_{\nu_\alpha}}{\mu_\nu} \frac{\delta \rho_{\nu_\alpha}}{\delta t} - \del{\rho_{\nu_\alpha}}{\mu_\nu} \frac{\delta n_{\nu_\alpha}}{\delta t} \biggr]~,\label{nu1}
\\ \frac{dT_\nu}{dt}&=  \biggl( \frac{\partial n_{\bar{\nu}_\alpha}}{\partial \mu_{\nu}} \frac{\partial \rho_{\bar{\nu}_\alpha}}{\partial T_{\nu}}-\frac{\partial n_{\bar{\nu}_\alpha}}{\partial T_{\nu}} \frac{\partial \rho_{\bar{\nu}_\alpha}}{\partial \mu_{\nu}} \biggr)^{-1}   \biggl[ -3H \biggl( (\rho_{\bar{\nu}_\alpha}+P_{\bar{\nu}_\alpha})\del{n_{\bar{\nu}_\alpha}}{\mu_\nu}-n_{\bar{\nu}_\alpha} \del{\rho_{\bar{\nu}_\alpha}}{\mu_\nu} \biggr) +\del{n_{\bar{\nu}_\alpha}}{\mu_\nu} \frac{\delta \rho_{\bar{\nu}_\alpha}}{\delta t} - \del{\rho_{\bar{\nu}_\alpha}}{\mu_\nu} \frac{\delta n_{\bar{\nu}_\alpha}}{\delta t} \biggr] .
\label{nu2}
\end{align}
%
In addition, each thermodynamic quantity for $\{\nu_\alpha \}, \;\{\bar{\nu}_\alpha \}$ is expressed by the particle number density $n_\nu$, the energy density $\rho_\nu$, and the pressure $P_\nu$ for the total neutrino~:
\begin{align}
n_{\nu_\alpha}&=n_{\bar{\nu}_\alpha}=\frac{1}{6}n_\nu~, 
\\ \rho_{\nu_\alpha} &= \rho_{\bar{\nu}_\alpha}=\frac{1}{6} \rho_\nu~,
\\ P_{\nu_\alpha}&=P_{\bar{\nu}_\alpha} =\frac{1}{6} P_\nu . 
\end{align}
By adding both sides of Eqs.~(\ref{nu1}) and (\ref{nu2}), and summing over all flavors, we obtain the evolution equation for $T_\nu$ (\ref{dTnu-4}).
The evolution equation for $\mu_\nu$ (\ref{dmunu-4}) can also be obtained in the same way.

For the Majoron evolution equation, by using  eqs.(\ref{dTa}) and (\ref{dmua}) set to $a=\phi$, we obtain the evolution equations for $T_\phi$ (\ref{dTphi}) and $\mu_\phi$ (\ref{dmuphi}).

\newpage
{\small 
\bibliographystyle{JHEP}
\bibliography{ref}

}

\end{document}